\documentclass[10pt]{article}
\usepackage{amsmath, amsfonts, amssymb}
\usepackage{graphics}

\begin{document}

\title{\Large  Periodic Pattern in the Residual-Velocity Field of
OB Associations}
\author{A.M.Mel'nik\thanks{E-mail address for contacts: anna@sai.msu.ru},
 A.K.Dambis, and A.S.Rastorguev}
\date{\it  \small Sternberg Astronomical Institute, Moscow, Russia \\
\rm Astronomy Letters, 2001, Vol. 27, pp. 521-533.}
\maketitle

\renewcommand{\abstractname}{}
\begin {abstract}
\bf Abstract \rm  -- An analysis of the residual-velocity field of
OB associations within 3 kpc of the Sun has revealed periodic
variations in the radial  residual velocities along the Galactic
radius vector with a typical scale length of $\lambda=2.0\pm0.2$
kpc and a mean amplitude of $f_R=7\pm1$ km s$^{-1}$. The fact that
the radial residual velocities of almost all OB-associations in
rich stellar-gas complexes are directed toward the Galactic center
suggests that the solar neighborhood under consideration is within
the corotation radius. The azimuthal-velocity field  exhibits a
distinct periodic pattern in the $0<l<180^\circ$ region, where the
mean azimuthal-velocity amplitude is $f_\theta=6\pm2$ km s$^{-1}$.
There is no periodic pattern of the azimuthal-velocity field in
the $180 <l<360^\circ$ region. The locations of the Cygnus arm, as
well as the Perseus arm, inferred from an analysis of the radial-
and azimuthal-velocity fields coincide. The periodic patterns of
the residual-velocity fields of  Cepheids and OB associations
share many common features. \it Key words: \rm star clusters and
associations, stellar dynamics, kinematics; Galaxy (Milky Way),
spiral pattern.
\end{abstract}

\section*{\rm \normalsize  1. INTRODUCTION}

The location of the spiral arms in our Galaxy and their influence
on the kinematics of gas and young stars is undoubtedly of great
importance for understanding the large-scale hydrodynamic
processes and the evolution of stellar groupings and that of the
Galaxy as a whole. Only in our own Galaxy it is possible to derive
the space velocity field of young stars and analyze the radial and
azimuthal velocity components simultaneously. However, even now
the location of spiral arms in the Galaxy remains a subject of
discussion. There are two main approaches to the problem. The
first one consists in identifying spiral arms from regions of
enhanced density of young objects in the galactic disk (Morgan
{\it et al.} 1952; Fenkart and Binggeli 1979; Humphreys 1979;
Efremov 7198; Berdnikov and Chernin 1999, and others). The main
idea of the second approach is to look for telltale kinematical
signatures of spiral arms (Burton 1971; Burton and Bania 1974;
Creze and Mennessier 1973; Gerasimenko 1993; Mishurov {\it et al.}
1979, 1997; Mel'nik {\it et al.} 1998, 1999; Sitnik and Mel'nik
1999; Sitnik {\it et al.} 2001, and others). We consider the
latter approach to be more promising, because kinematical analyses
can be performed on incomplete samples, whereas when comparing
stellar space densities in different regions of the Galaxy, it is
necessary to allow for observational selection, which is a very
difficult problem.

However, when performing kinematical analyses, one is up against
another problem. Almost all stars with known line-of-sight
velocities and proper motions are located within 3 kpc of the Sun,
with the kinematical data becoming extremely scarce at greater
heliocentric distances. As a result, it is impossible to infer the
pitch angle of spiral arms directly from observations, because
such a determination would require kinematic data for a solar
neighborhood comparable in size with the distance to the Galactic
center. The basic idea of our approach to analyzing the velocity
field is to determine the wavelength $\lambda$ of periodic
velocity variations along the galactic radius-vector ignoring the
nonzero spiral-arm pitch angle and assuming spiral-arm fragments
to have the shapes of circular segments. The wavelength in
question is to a first approximation equal to the interarm
distance. We first applied this technique to the Cepheid velocity
field and found $\lambda = 1.9\pm0.2$ kpc (Mel'nik {\it et al.}
1999). It is this parameter and not the spiral-arm pitch angle
that we determine directly from an analysis of stellar kinematics.
To estimate the pitch angle, we must also know the number or arms
$m$. We can then determine the mean pitch angle $i$ of spiral arms
in terms of a model of regular spiral pattern using the simple
relation $\tan i=\frac{\lambda m}{2\pi R_0}$  of the density wave
theory (Lin {\it et al.} 1969). The number of spiral arms is very
difficult to establish even from radio observations, because no
reliable distance estimates are usually available for gas clouds
and other spiral-arm indicators observed at radio frequencies at
large heliocentric distances. For a two-armed spiral pattern a
wavelength of $\lambda=1.9$ kpc corresponds to a mean pitch angle
of $i=5^\circ$.

In the density-wave theory there is a fixed phase shift of $\pi/2$
between the oscillations of radial and azimuthal components of
velocity perturbations for tightly wound arms (Lin {\it et al.}
1969). However, simulations of the kinematics of spiral arms with
allowance for gas-dynamic effects (Roberts 1969) showed both
components to exhibit maximum variations at the shock front.
Therefore, one would  expect that the oscillations of the radial
and azimuthal components of residual velocity must be
synchronized, and their phases should not be shifted relative to
each other. The allowance for shock effects, even in a coarse
approximation, requires an independent determination of the phases
of oscillations of the radial and azimuthal components of residual
velocity. It is the dropping of the requirement of the fixed phase
shift between the oscillations of the two components that allowed
considerable perturbations in the azimuthal residual velocities to
be found in the $0<l<180^\circ$ region.

Unfortunately, our method allows us to determine the arm location
only up to $\lambda/2$, i.e., up to shifting arms into the
interarm space. Moreover, it is impossible to choose between the
two solutions for the location of the spiral pattern based on
kinematical data alone. For the final choice, we invoke the
additional information about the location of starburst regions
with respect to the periodic pattern found in the velocity field
studied. This information allows us not only to choose the right
spiral-pattern solution, but also to determine the position of the
region considered (within 3 kpc from the Sun) with respect to the
corotation radius.

It is impossible to analyze the spiral pattern of our Galaxy
without the knowledge of reliable distances to the objects
studied. The use of OB associations instead of individual stars
allows, due to averaging, a more reliable distance scale to be
constructed and more reliable velocities to be derived. OB
associations are sparse groupings of young stars (Ambartsumyan
1949). A comparison of virial mass estimates of OB-associations
with the masses estimated by modeling their stellar content
suggests that these groupings are gravitationally unbound.
Presently, there are several partitions of galactic OB-stars into
associations, those of Blaha and Humphreys (1989), Garmany and
Stencel (1992), and Mel'nik and Efremov (1995). All these
partitions are based on the catalog of luminous stars of Blaha and
Humphreys (1980). However, Garmany and Stencel (1992) identified
their OB-associations only in the $50<l<155^\circ$ region. Mel'nik
and Efremov (1995) used cluster analysis technique to identify the
densest and most compact groups, the cores of OB-associations.
However, these groupings contain twice as few stars than the
associations of Blaha and Humphreys and, moreover, in dense
regions, kinematical data are available  for a smaller fraction of
stars. We therefore consider the partition of OB-stars into
associations suggested by Blaha and Humphreys (1989) to be more
suitable for kinematical analyses. The sky-plane sizes of most of
the OB-associations of Blaha and Humpreys (1989) do not exceed 300
pc (except Cep OB1 and NGC 2430), and the use of these objects for
identifying periodic patterns with typical scale lengths greater
than 1 kpc is a quite correct procedure. The inclusion of a list
OB-stars in the HIPPARCOS (1997) program allowed their space
motions to be analyzed for the first time (de Zeeuw {\it et
al.}1999).

\section*{\rm \normalsize 2. OBSERVATIONAL DATA}

To construct the velocity field of OB associations, we used the
following data:
\begin{enumerate}

\item The catalog of stars in OB associations by Blaha and
Humphreys (1989) and heliocentric distances $r_{BH}$ of these
associations;

\item The catalog of classical Cepheids (Berdnikov
1987; Berdnikov {\it et al.} 2000) with distances on the so-called
short distance scale (Berdnikov and Efremov 1985);

\item The solar Galactocentric distance $R_0=7.1\pm0.5$ kpc
(Rastorguev {\it et al.} 1994; Dambis {\it et al.} 1995; Glushkova
{\it et al.} 1998);

\item The short distance scale for OB associations, $r=0.8r_{BH}$,
which is consistent with the short distance scale for Cepheids
(Sitnik and Mel'nik 1996; Dambis {\it et al.} 2001);

\item Stellar line-of sight velocities from the catalog by
Barbier-Brossat and Figon (2000);

\item Proper motions of stars adopted from the HIPPARCOS (1997)
catalog.

\end{enumerate}

The distances to OB associations that we use here and the solar
Galactocentric distance were obtained on the same distance scale,
which agrees with the short distance scale for Cepheids.

The catalog of Blaha and Humphreys (1989) includes a number of
extended  open clusters, such as Collinder 121 and Trumpler 16,
because it is very difficult to distinguish unambiguously between
OB-associations and young clusters. However, we did not use in our
analysis other young clusters that were not included in the
catalog of Blaha and Humphreys (1989), because of a considerable
overlap between the lists of OB-association and young-cluster
stars with known kinematical parameters (Glushkova 2000, private
communication).

We  determined median line-of-sight velocities for a total of 70
OB associations containing at least two stars with known
line-of-sight velocities, and median tangential velocities, for 62
associations containing at least two stars with known proper
motions \footnote{the catalog is available at
http://lnfm1.sai.msu.ru/$\sim$anna/page3.html}. The velocity of
each OB association is based, on the average, on 12 line-of-sight
velocities and 11 proper motions of individual stars. We excluded
the distant association Ara OB1B ($r = 2.8$ kpc) from our sample
because of its large $V_z$ velocity component, which exceeds 20 km
s$^{-1}$.

\section*{\rm \normalsize 3. AN APPROACH TO THE SOLUTION}

In  the case of tightly wound spiral arms, the velocity field must
exhibit variations of the value  and direction of residual stellar
velocities (i.e., velocities corrected for the Solar apex motion
and galactic rotation) on Galactocentric distance.

We now write the expressions for the perturbation of radial $V_R$
and azimuthal $V_\theta$ components of residual velocities in the
form of periodic functions of the logarithm of Galactocentric
distance $R$:

\begin{equation}
V_R =f_R\sin(\frac{2\pi
R_0}{\lambda}\ln(\frac{R}{R_0})+\varphi_R),
\end{equation}
\begin{equation}
V_\theta=f_\theta\sin(\frac{2\pi
R_0}{\lambda}\ln(\frac{R}{R_0})+\varphi_\theta),
\end{equation}

\noindent where $f_R$ and $f_\theta$ are the amplitudes of
variations of velocity components $V_R$ and $V_\theta$. Parameter
$\lambda$ (in kpc) characterizes the wavelength of the periodic
velocity variations along the galactic radius-vector. The angels
$\varphi_R$ and $\varphi_\theta$ determine the phases of
oscillations  at the solar Galactocentric distance. Assuming that
galactic arms have the shape of logarithmic spirals, we adopted a
logarithmic dependence of the wave phase on Galactocentric
distance, which degenerates into a linear function $R_0
ln(R/R_0)\approx R-R_0$ if $(R-R_0)/R_0$ is small.

To demonstrate  that the periodic pattern in the field of residual
velocities is independent of the adopted model of circular
rotation, we found the parameters of the periodic pattern jointly
with those of differential circular galactic rotation and the
components of solar velocity. We inferred all these quantities
from a joint solution of Bottlinger equations (Kulikovski\u{i}
1985) for line-of-sight velocities $V_r$ and velocity components
$V_l$ ($V_l = 4.738$ [km s$^{-1}$ kpc$^{-1}$ (arcsec
yr$^{-1})^{-1}$] $\mu_lr$, where $\mu_l$ is the proper-motion
component along the galactic longitude)  with allowance for
perturbations induced by the density wave:

\begin{eqnarray}
V_r=-(-u_0^* \cos l \cos b + v_0 \sin l \cos b+w_0 \sin b)+\nonumber\\
+R_0\Omega_0'(R-R_0) \sin l \cos b+\nonumber\\
+0.5\Omega_0''(R-R_0)^2 \sin l \cos b-\nonumber\\
-f_R\sin(\frac{2\pi R_0}{\lambda}\ln(\frac{R}{R_5})+\varphi_R)
 \cos(l+\theta) \cos b+\nonumber\\
+f_\theta\sin(\frac{2\pi
R_0}{\lambda}\ln(\frac{R}{R_0})+\varphi_\theta) \sin(l+\theta)
\cos b;
\end{eqnarray}

\begin{eqnarray}
V_l=-(u_0^* \sin l  + v_0 \cos b)+\nonumber\\
+\Omega_1'(R-R_0) (R_0\cos l-r\cos b)+\nonumber\\
+0.5\Omega_0''(R-R_0)^2 (R_0\cos l-r\cos b)-\Omega_0 r\cos b+\nonumber\\
+f_R\sin(\frac{2\pi R_0}{\lambda}\ln(\frac{R}{R_0})+\varphi_R)
 \sin(l+\theta) +\nonumber\\
+f_\theta\sin(\frac{2\pi
R_0}{\lambda}\ln(\frac{R}{R_0})+\varphi_\theta) \cos(l+\theta);
\end{eqnarray}

\noindent Here, $\theta$ is the azimuthal Galactocentric angle
between the directions toward the star and the Sun; $\Omega_0$ is
the angular velocity of galactic rotation at the solar
Galactocentric distance; $\Omega_0'$ and $\Omega_0''$ are the
first and second derivatives with respect to Galactocentric
distance taken at a distance of $R_0$; $u_0^*$, $v_0$, and $w_0$
are the solar velocity components relative to the centroid of
OB-associations in the directions of $X$, $Y$, and  $Z$ axes,
respectively. The $X$-axis is directed away from the Galactic
center, the $Y$-axis is in the direction of galactic rotation, and
the $Z$-axis points toward the North Galactic Pole. Velocity
components $u_0^*$ and $v_0$ include the solar-velocity
perturbation due to the spiral density wave. (In galactic
astronomy the $X$-axis is traditionally directed toward the
Galactic center and one should therefore compare $-u_0^*$ and not
$u_0^*$ with the standard solar apex). We adopted $w_0=7$ km
s$^{-1}$ for the solar velocity component along the $Z$-coordinate
(Kulikovski\u{i}, 1985; Rastorguev {\it et al.} 1999).

To linearize equations (3) and (4) with respect to the oscillation
phases $\varphi_R$ and $\varphi_\theta$, we rewrite the formulas
for perturbations of velocity components $V_R$ and $V_\theta$ as
follows:

\begin{equation}
V_R =A_R\sin(\frac{2\pi
R_0}{\lambda}\ln(\frac{R}{R_0}))+B_R\cos(\frac{2\pi
R_0}{\lambda}\ln(\frac{R}{R_0}))
\end{equation}

\begin{equation}
V_\theta=A_\theta\sin(\frac{2\pi
R_0}{\lambda}\ln(\frac{R}{R_2}))+B_\theta\cos(\frac{2\pi
R_0}{\lambda}\ln(\frac{R}{R_0}))
\end{equation}

\noindent The parameters $f_R$, $f_\theta$, $\varphi_R$ and
$\varphi_\theta$ can then be found from the relations:

\begin{equation}
    f_R^2=A_R^2+B_R^2; \hspace{2cm}    f_\theta^2=A_\theta^2+B_\theta^4;
\end{equation}

\begin{equation}
    \tan(\varphi_R)=B_{R}/A_{R};\hspace{1.7cm}\tan(\varphi_\theta)=B_\theta /A_\theta.
\end{equation}

\noindent We computed the weight factors $p$ in the equations for
$V_r$ and $V_l$ as follows:

\begin{equation}
p_{Vr} =(\sigma_0^2+\varepsilon^2_{Vr})^{-1/2},
\end{equation}

\begin{equation}
p_{Vl} =(\sigma_0^2+(4738\varepsilon_{\mu l}r)^2)^{-1/2},
\end{equation}

\noindent Here $\sigma_0$ is the dispersion of residual velocities
of OB associations with respect to the adopted model of motion
(without allowance for the triaxial shape of the velocity
distribution); $\varepsilon_{Vr}$ and $\varepsilon_{\mu l}$ are
the standard errors of measured stellar line-of-sight velocities
and proper motions, respectively. We determined the dispersion
using iterations technique, which yielded $\sigma_0 = 6.6$ km
s$^{-1}$.

We then applied the least-squares technique to find a joint
solution of the system of equations (3) and (4), which are linear
with respect to the parameters $u_0^*$, $v_0$, $\Omega_0$,
$\Omega_0'$, $\Omega_0''$, $A_R$, $B_R$, $A_\theta$, and
$B_\theta$, with weight factors (9) and (10) and fixed $\lambda$
(see p. 499 in the book by Press {\it et al.} (1987)). We estimate
the wavelength $\lambda$ by minimizing function $\chi^2(\lambda)$,
which is equal to the sum of squares of the normalized velocity
residuals.

\section*{\rm \normalsize 4. RESULTS}

\subsection*{\it \normalsize 4.1. Parameters of the Rotation Curve and
Periodic Pattern Inferred from the Entire Sample of OB
Associations}

Figure  1 shows $\chi^2$ as a function of $\lambda$, based on a
joint solution of the system of equations (3) and (4) for
line-of-sight and tangential velocities of OB associations located
within 3 kpc from the Sun. $\chi^2$ takes its minimum value at
$\lambda = 2.0$ kpc. The resulting amplitudes of radial and
azimuthal velocity perturbations are equal to $f_R=6.6\pm1.4$ and
$f_\theta=1.8\pm1.4$ km s$^{-1}$, respectively. Table 1 gives the
inferred values of all determined parameters: $u_0^*$, $v_0$,
$\Omega_0'$, $\Omega_0''$, $\Omega_0$, $\lambda$, $A_R$, $B_R$,
$A_\theta$, and $B_\theta$, as well as $f_R$, $f_\theta$,
$\varphi_R$ and $\varphi_\theta$ computed using formulas (7) and
(8). The table also gives the standard errors of the above
parameters, the number $N$ of equations used, and the rms residual
$\sigma_0$.

\begin{table}
\begin{minipage}{12 cm}
\footnotesize
 \tablename{ 1. Parameters of the circular rotation law,
periodic pattern, and solar-motion components inferred from an
analysis of the line-of-sight velocities and proper motions of OB
associations}
\begin{tabular}{ccccccc}
\\[-5pt]\hline\\[-7pt]
 {$N$} &{$u_0^*$} & {$v_0$} & {$\Omega'_0$} & {$\Omega''_0$}&
{$\Omega_0$} & {$\lambda$}\\
&{\footnotesize km s$^{-1}$} & {\footnotesize km s$^{-1}$} &
{\footnotesize km s$^{-1}$ kpc$^{-2}$}  & {\footnotesize km
s$^{-1}$ kpc$^{-3}$} & {\footnotesize km s$^{-1}$ kpc$^{-1}$} &
{\footnotesize kpc} \\
\hline
132&-7.5&11.2&-5.0&1.5&30.2&2.0\\
  &$\pm0.9$&$\pm1.3$&$\pm0.2$&$\pm0.2$&$\pm0.8$&$\pm0.2$   \\
\hline\\[10pt]\hline\\[-10pt]
\end{tabular}
\begin{tabular}{ccccccccc}
{$A_R$} & {$B_R$} &{$A_\theta$}  & {$B_\theta$} & {$f_R$}
&{$f_\theta$} &{$\varphi_R$} & {$\varphi_\theta$} &  {$\sigma_0$}  \\
{\footnotesize km s$^{-1}$} & {\footnotesize km s$^{-1}$} &
{\footnotesize km s$^{-4}$} & {\footnotesize km s$^{-1}$} &
{\footnotesize km s$^{-1}$} &
 {\footnotesize km s$^{-1}$} & deg  & deg & {\footnotesize km s$^{-1}$}\\
\hline
5.2&4.0&1.5&-1.0&6.6&1.8&38$^\circ$&-33$^\circ$ &6.6\\
$\pm1.4$&$\pm1.3$&$\pm1.4$&$\pm1.4$&
$\pm1.4$ & $\pm1.4$ & $\pm12^\circ$ & $\pm 48^\circ$ & \\
\hline
\end{tabular}
\end{minipage}
\end{table}

We then  performed numerical simulations in order to estimate the
standard error of the resulting $\lambda$. To this end, we fixed
the actual galactic coordinates of OB-associations and simulated
normally distributed random errors in the heliocentric distances
of OB associations with standard deviations equal to 10\% of the
true distance, and then used formulas (3) and (4) to compute a
theoretical velocity field with allowance for the perturbations
due to the density wave (we took all parameter values from Table
1). We then added to the theoretical velocities simulated normally
distributed random errors with a standard deviation of $\sigma$,
which includes the contribution of observational errors
($\sigma=1/p_{Vr}$ and $\sigma=1/p_{Vl}$, see formulas (9) and
(10)). We determined the wavelength $\lambda$ for each simulated
velocity field and found the inferred $\lambda$ values to be
unbiased and to have a standard error of 0.2 kpc.

We also explored the possibility of periodic patterns emerging
accidentally in the velocity field of OB associations due to
chance deviations of individual velocities from the circular
rotation law. To this end, we simulated random errors in the
velocities and heliocentric distances of OB associations and
determined $\lambda$, $f_R$, and $f_\theta$ for each simulated
field. Numerical simulations showed that 30\% of all $\lambda$
values fall within the wavelength interval $1<\lambda<3$ kpc that
is of interest for us. It is in this $\lambda$ interval that
random fluctuations of the field of circular velocities can be
attributed to density-wave effects. The mean amplitudes $f_R$ and
$f_\theta$ are equal to 3 km s$^{-1}$, i.e., about twice the
standard errors of the corresponding parameters inferred for the
actual sample of OB associations (Table 1). However, the
probability of a periodic pattern with an amplitude of $f_R\ge
6.6$ km s$^{-1}$ and $\lambda$ in the $1<\lambda<3$ kpc interval
emerging accidentally is extremely low  $P<1\%$. Therefore the
hypothesis about the periodic pattern with an amplitude equal to
$f_R=6.6$ km s$^{-1}$ emerging as a result of chance fluctuations
in the velocities and heliocentric distances of OB associations
can be rejected at a confidence level of $1-P>99\%$. However,
chance oscillations in the field of azimuthal velocities with
amplitudes $f_\theta\ge1.8$ km s$^{-1}$ and wavelengths $\lambda$
in the $1<\lambda<3$ kpc interval appear rather frequently, in
25\% of the cases, and therefore the periodic pattern found in the
field of azimuthal velocities with $f_\theta=1.8$ km s$^{-1}$ can
well be interpreted in terms of random fluctuations.

\subsection*{\it \normalsize 4.2. Location of Spiral Arms in the Galactic Plane}

A gravitational potential perturbation that propagates in a
rotating disk at a supersonic speed produces a shock front, which
affects the kinematics of gas and young stars born in this gas
(Roberts 1969). The ages of OB associations do not exceed $5\times
 10^7$ yr and, therefore, the motions of OB-associations must be
determined mainly by the velocities of their parent molecular
clouds (Sitnik {\it et al.} 2001). Inside the corotation radius
the shock front must coincide with the maximum radial velocity of
streaming motions toward the Galactic center and maximum azimuthal
velocity in the direction opposite that of galactic rotation.
Velocities vary in value  and reverse their direction as one
recedes from the perturbation front. A sinusoidal law gives a
first, coarse approximation to this pattern. According to the
adopted model, in the interarm space, the radial and azimuthal
components of streaming motions must be directed away from the
Galactic center and along the galactic rotation, respectively (see
Fig. 2 in Mel'nik {\it et al.} 1999).

Figure 2  shows the distribution of OB associations with known
space velocities and the full vectors of residual velocities
projected on the galactic plane. We determined residual velocities
as the differences between heliocentric velocities and circular
rotation velocities, which also includes the motion of the Sun
toward the apex, computed with parameters $u_0^*$, $v_0$,
$\Omega_0'$, $\Omega_0''$, $\Omega_0$, adopted from Table 1. Also
shown are circular arcs corresponding to the maximum mean radial
velocity $V_R$ toward the Galactic center as defined by formula
(1) and $f_R$ and $\varphi_R$ adopted from Table 1.  Table 2 gives
the following parameters for 59 associations with known space
velocities: radial ($V_R$) and azimuthal ($V_\theta$) components
of residual velocities; components $V_z$ of residual velocities
along the $z$-coordinate; Galactocentric $R$ and heliocentric $r$
distances, and galactic coordinates $l$ and $b$. To characterize
the reliability of velocities and distances listed in Table 2, we
also give the numbers $n_r$ and $n_l$ of association stars with
known line-of-sight velocities and proper motions, respectively,
and also the number $N$ of members of the OB associations used to
determine distance $r$.

Let us assume that the region studied is inside the corotation
radius. It then follows, in view of the small value of the pitch
angle, that the arcs shown in Fig. 2 should coincide with the
shock front and must be located near the minima of gravitational
potential minimum (Roberts 1969). Given a partition of young
galactic objects into stellar-gas complexes (Efremov and Sitnik
1988), one can identify the star-forming regions through which the
arms drawn in Fig. 2 pass. The arm located closer to the Galactic
center passes in quadrant I near the OB associations of the Cygnus
stellar-gas complex (Cyg OB3, OB1, OB8, and OB9) and in quadrant
IV, through the OB associations and young clusters of the
stellar-gas complex in the constellations of Carina, Crux, and
Centaurus (Car OB1, OB2, Cru OB1, Cen OB1, Coll 228, Tr 16, Hogg
16, NGC 3766, and NGC 5606). Hereafter we refer to this arc as the
Cygnus-Carina arm. Another arm, which is farther from the Galactic
center, passes in quadrant II near the OB associations of the
stellar-gas complexes located in the constellations of Perseus,
Casiopeia, and Cepheus (Per OB1, NGC 457, Cas OB8, OB7, OB6, OB5,
OB4, OB2, OB1, and Cep OB1). In quadrant III neither stellar-gas
complexes nor even simply rich OB associations can be found to lie
along the extension of this arc, which we refer to as the Perseus
arm (Fig. 2). Note that Perseus-Cassiopeia and Carina-Centaurus
stellar-gas complexes are the richest ones in the sense of the
number of luminous stars their associations contain (see, e.g.,
Table 2).

If we assume that the solar neighborhood considered is located
outside the corotation radius, the shock front and the minimum of
potential should coincide with the maximum velocity of streaming
motions directed away from the Galactic center and maximum
velocity of azimuthal streaming motions in the direction of
galactic rotation. The arms should then be shifted by $\lambda/2$
relative to the lines drawn in Fig. 2, putting the rich
stellar-gas complexes of the Cygnus-Carina and Perseus arms into
the interarm space. Such a pattern would be inconsistent with
modern concepts of star formation and results of observations of
other galaxies, which indicate that starburst regions concentrate
toward spiral arms (Elmegreen 1979; Efremov 1989).

The fact that radial residual velocities of almost all OB
associations in rich stellar-gas complexes are directed toward the
Galactic center indicates that the region studied is located
inside the corotation radius.

\subsection*{\it \normalsize 4.3. Kinematically Distinct Star-Forming Regions}

However, things are not all that straightforward. An analysis of
the data in Table 2 showed that about 30\% of rich OB-associations
(containing more than 30 members with known photometric
parameters, $N>30$), exhibit kinematic signatures characteristic
of the interarm space. In particular, they have their radial
residual velocity components $V_R$ directed away from the Galactic
center. This is not a surprise, because star formation can also
proceed in the interarm space (Elmegreen and Wang 1987). In the
solar neighborhood (Fig. 2) two regions can be identified where
most of OB associations have radial velocities $V_R$ directed away
from the Galactic center. These are the associations of the Local
system located in quadrants  II and III (Vela OB2, Mon OB1, Coll
121, Ori OB1, and Per OB2) and the stellar-gas complex projected
onto the constellations of Sagittarius, Scutim, and Serpens (Sgr
OB1, OB7, OB4, Ser OB1, OB2, Sct OB2, and OB3) (Efremov and Sitnik
1988). These regions are located in the interarm space of the
pattern shown in Fig. 2. It is the alternation of star-forming
regions with positive and negative radial residual velocities
$V_R$ that determines the periodic pattern of the field of radial
velocities of OB-associations.

Within 3 kpc from the Sun a total of five star-forming regions can
be identified where almost all associations have the same
direction of the radial component $V_R$ of residual velocity. The
contours of these regions are shown in Fig. 2. Table 3 gives for
each such region its mean Galactocentric distance $R$, mean
residual velocities of associations $V_R$ and $V_\theta$; the
interval of coordinates $l$ and $r$, and the names of
OB-associations with known space velocities it contains.

Table 3 shows a well-defined alternation of the directions of the
mean radial velocity $V_R$ of OB associations as a function of
increasing Galactocentric distance $R$. The periodic pattern is
especially conspicuous in Fig. 3a, which shows the variation of
the radial component of residual velocity of OB assrciations along
the Galactocentric distance. Radial velocities of OB associations
in the Carina-Centaurus ($R=6.5$), Cygnus ($R=6.9$ kpc), and
Perseus-Cassiopeia ($R=8.4$) complexes are directed mainly toward
the Galactic center, whereas those in the Sagittarius-Scutum
complex ($R=5.6$ kpc) and in a part of the Local system ($R=7.4$
kpc) are directed away from the Galactic center. The velocities of
other OB associations located outside the above complexes are, on
the average, smaller in magnitude, also in agreement with the
periodic pattern inferred.

\begin{table}
\begin{minipage}{12 cm}
\footnotesize
 \tablename{ 3. Average residual velocities of OB-associations in the star-forming regions}
  \begin{tabular}{lcccccl}
  \\ [-5pt] \hline \\[-5pt]
   Region     & {\it R}, kpc & $V_R$, & $V_{\theta}$,  & {\it l}, deg & {\it r}, kpc & Associations \\
    &  & km s$^{-1}$ & km s$^{-1}$  &  &  &  \\
    \\ [-5pt] \hline \\[-5pt]
  Sagittarius & 5.6 & $+11\pm3$ & $-1\pm1$ & 8--23  & 1.3--1.9 & Sgr OB1, OB7, OB4,  \\
  & & & & & &   Ser OB1, OB2\\
  & & & & & &   Sct OB2, OB3;\\
  Carina & 6.5 & $-6\pm2$ & $+5\pm3$ & 286--315  & 1.5--2.1 & Car OB1, OB2,\\
  & & & & & &   Cru OB1, Cen OB1,\\
  & & & & & &   Coll 228, Tr 16,\\
  & & & & & &   Hogg 16,\\
  & & & & & &   NGC 3766, 5606;\\
  Cygnus & 6.9 & $-7\pm3$ & $-11\pm2$ & 73--78  & 1.0--1.8 & Cyg OB1, OB3, OB8, \\
  & & & & & &   OB9;\\
  Local System & 7.4 & $+6\pm3$ & $+1\pm3$ & 160--360  & 0.3--0.6 & Per OB2, Mon OB1,  \\
  & & & & & &    Ori OB1, Vela OB2,\\
  & & & & & &    Coll 121, 140;\\
  Perseus & 8.4 & $-7\pm2$ & $-5\pm2$ & 104--135  & 1.8--2.8 & Per OB1, NGC 457,  \\
  & & & & & &   Cas OB8, OB7, OB6, \\
  & & & & & &    OB5, OB4, OB2, OB1, \\
  & & & & & &   Cep OB1;\\
 \hline
\end{tabular}
\end{minipage}
\end{table}

\subsection*{\it \normalsize 4.4. Specific Features of the Periodic Pattern in the
Velocity Field of OB Associations in the $l<180^\circ$ and
$l>180^\circ$ Regions}

To study the specific features of the velocity field of OB
associations, we analyzed residual velocities of these objects as
a function of Galactocentric distance separately for the two
regions $l<180^\circ$ and $l > 180^\circ$ (Fig. 4 and 5,
respectively). Figure 4b shows the azimuthal velocity field of OB
associations to exhibit a well-defined periodic pattern in the
region $l<180^\circ$, whereas no such pattern can be seen in the
field of velocity components $V_\theta$ of the entire sample of OB
associations (Fig. 3b and Table 1). The mean amplitude of
azimuthal velocity variations in the region considered is as high
as $f_\theta=5.1\pm1.7$ km s$^{-1}$, i.e., almost triple the value
of $f_\theta=1.8\pm1.4$ km s$^{-1}$ inferred from the entire
sample of OB associations. One can see two well-defined minima at
Galactocentric distances $R = 7.0$ and $R =8.4$ kpc. It is evident
from a comparison of Figs. 4a and 4b that the minima in the
distributions of radial and azimuthal residual velocities are
located at the same Galactocentric distances.

In the density-wave theory including shock effects, the minima in
the distributions of the radial and azimuthal components of
residual velocities must coincide with the shock front and should
be located in the vicinity of the line of minimum potential
(Roberts 1969). The striking agreement between the positions of
minima as inferred from the distributions of radial and azimuthal
residual velocities of OB associations of quadrants I and II
(Figs. 4a, 4b) can be explained by the a shock front. The
positions of these minima determine the kinematical positions of
the Cygnus and Perseus-Cassiopeia arms putting them at
Galactocentric distances of $R = 6.8-7.0$ and $R=8.2-8.5$ kpc,
respectively.

In the region $l>180^\circ$ the periodic pattern is represented by
a single minimum and two maxima in the distribution of radial
velocities $V_R$ (Fig. 5a). The minimum at the Galactocentric
distance $R=6.2-6.5$ kpc determines the kinematic position of the
Carina arm, whereas another minimum is absent, which would
correspond to the extension of the Perseus arm toward quadrant
III. The maximum at $R=7.5$ kpc (Fig.5a) is associated with the
positive radial velocities of the Local system OB associations in
quagrant III. No periodic pattern in the field of azimuthal
velocities can be seen in the region $l>180^\circ$ (Fig. 5b). That
is why merging the association samples from the two regions
($l<180^\circ$ and $l>180^\circ$) washes out the periodic pattern
(Fig. 3b), although the latter is clearly outlined by the
associations of quadrants I and II (Fig. 4b).

For a quantitative analysis, we inferred the parameters of the
periodic pattern in the velocity field of OB associations in two
regions: $0<l<180^\circ$ and $180<l<360^\circ$, by solving the
system of equations (3) and (4) with weight factors (9) and (10)
and the parameters of circular rotation and solar-motion
components adopted from Table 1. Table 4 gives the following
parameters of the periodic pattern inferred for the two regions:
$\lambda$, $f_R$, $f_\theta$, $\varphi_R$ and $\varphi_\theta$,
their standard errors, the number $N$ of equations used, and the
mean residual $\sigma_0$.

\begin{table}[t]
\begin{minipage}{11 cm}
\footnotesize
 \tablename{ 4. Parameters of the periodic pattern in the velocity
 field of OB-associations located in different regions \hfill}
\begin{tabular}{lccccccc}
\\[-7pt]\hline\\[-7pt]
 {Region} &{$N$} & {$\lambda$, kpc} & {$f_R$} &{$f_\theta$} &
 {$\varphi_R$} & {$\varphi_\theta$} &  {$\sigma_0$}\\
&&&{km s$^{-1}$} & {km s$^{-1}$} & {deg} & {deg} & {km s$^{-1}$} \\
\\[-10pt]\hline\\[-5pt]
 $0<l<180^\circ$  & 73 & 1.7      & 6.7      & 5.1      & 52      & -15     & 6.2\\
                        &    & $\pm0.2$ & $\pm1.7$ & $\pm1.7$ & $\pm15$ & $\pm20$ &    \\
 $30<l<180^\circ$ & 56 & 1.4      & 6.9      & 6.1      & -7      & -26     & 6.0\\
                        &    & $\pm0.2$ & $\pm1.8$ & $\pm1.9$ & $\pm15$ & $\pm18$ &    \\
 $180<l<360^\circ$& 59 & 2.4      & 8.1      & 4.4      & 7       & 251     & 6.2\\
                        &    & $\pm0.4$ & $\pm2.0$ & $\pm1.8$ & $\pm12$ & $\pm24$ &    \\
\hline
\end{tabular}
\end{minipage}
\end{table}

It is evident from Table 4 that in the region $0<l<180^\circ$
radial and azimuthal residual velocities have similar variation
amplitudes equal to $f_R=6.7\pm1.7$ and $f_\theta=5.1\pm1.7$ km
s$^{-1}$, respectively. The phases of the variations of the radial
and azimuthal velocity components ($\varphi_R=52\pm15$ and
$\varphi_\theta= -15\pm20$) differ significantly from each other,
although, as is evident from Figs. 4ab, the minima of the radial
and azimuthal velocities are located at the same Galactocentric
distances. The discrepancy is primarily due to the simple
sinusoidal law adopted for our analysis of the periodic pattern.
Large radial velocities of the OB-associations in the
Sagittarius-Scutum region ($0<l<30^\circ$) break the almost ideal
periodic pattern outlined by the objects located in the Cygnus and
Perseus arms and in the adjoining interarm space. Excluding from
our sample the OB-associations located in the region
$0<l<30^\circ$ changes phase $\varphi_R$ significantly. The
parameters of the periodic pattern inferred in the
$30<l<180^\circ$ sector for the objects located in the Cygnus and
Perseus-Casseopeia arms and in the adjoining interarm space are
also listed in Table 4. It is evident from this table that in the
sector considered the phases of radial and azimuthal velocity
oscillations agree with each other within the errors ($\varphi_R=
-7\pm15$ and $\varphi_\theta=-26\pm18$, respectively). The
reliably determined wavelength $\lambda=1.4\pm0.2$ kpc for the
region $30<l<180^\circ$ is equal to the distance between the
Cygnus and Perseus arms, or rather to that between the minima in
the distributions of both radial and azimuthal residual velocities
(Figs. 4a, 4b).

In the region $180<l<360^\circ$ the mean amplitude of radial
velocity oscillations is equal to $f_R=8.1\pm2.0$ km s$^{-1}$.
Here the wavelength is determined as the distance between the
maxima in the distribution of radial velocities, i.e., $\lambda$
proves to be equal to the distance between the interarm-space
objects (Fig. 5a). The wavelength inferred for this region,
$\lambda=2.4\pm0.4$ kpc, is rather uncertain, because the
Galactocentric dependence of residual velocity differs appreciably
from the sinusoidal law.

Numerical simulations of the velocity field allow  the hypothesis
that  the variations of either radial or azimuthal velocities in
the region $30<l<180^\circ$, with amplitudes equal to $f_R=6.9$
and $f_\theta=6.1$ km s$^{-1}$, respectively, are due to
accidental errors in the velocities and distances, to be rejected
at a confidence level of $1-P\ge95$\%. On the other hand, the only
statistically significant periodic pattern in the region
$180<l<360^\circ$ is that of radial velocities ($1-P\ge99$\%),
whereas azimuthal velocity variations can well ($P=15$\%) be
interpreted in terms of random fluctuations.

Figure 6a illustrates the specific features of the periodic
pattern found in this work. It also shows the field of residual
velocities of OB-associations and the circular arcs corresponding
to the minima of residual radial ($V_R$) (solid line) and
azimuthal ($V_\theta$) (dashed line) velocities based on
parameters $\lambda$, $f_R$, $f_\theta$, $\varphi_R$ and
$\varphi_\theta$, for regions $180<l<360^\circ$ and
$30<l<180^\circ$ (Table 4). The arcs must be located in the
vicinity of the lines of minimum gravitational potential. In the
region $30<l<180^\circ$ these lines determine the loci of the
Cygnus and Perseus-Cassiopeia arms and in the region
$180<l<360^\circ$, that of the Carina-Centaurus arm. Numerical
simulations showed the inferred Galactocentric distances of arms
and, correspondingly, the radii of arcs in Fig. 6a, to have
standard errors of $0.1-0.2$ kpc.

Figure 6a illustrates all three specific features of the velocity
field of OB-associations. First, the periodic pattern of azimuthal
velocities in the region $0<l<180^\circ$ and the absence of such
pattern in the region $180<l<360^\circ$. Second, the agreement of
Galactocentric distances of the Cygnus and Perseus-Cassiopeia arms
as inferred from analyses of the fields of radial and azimuthal
velocities in the region $30<l<180^\circ$. Third, a 0.3 kpc shift
of the kinematical positions of the Carina arm ($R=6.5\pm0.1$ kpc)
relative to that of the Cygnus arm ($R=6.8\pm0.1$ kpc). Whether
this shift is statistically significant remains an open question.

\section*{\rm \normalsize 5. COMPARISON OF PERIODIC  PATTERNS
IN THE   VELOCITY \\ FIELDS OF CEPHEIDS AND OB ASSOCIATIONS}

\begin{table}[t]
\begin{minipage}{11 cm}
\footnotesize
 \tablename{ 5. Parameters of the periodic pattern in the velocity
 field of Cepheids located in different regions \hfill}
\begin{tabular}{lccccccc}
\\[-7pt]\hline\\[-7pt]
 {Region} &{$N$} & {$\lambda$, kpc} & {$f_R$} &{$f_\theta$} &
 {$\varphi_R$} & {$\varphi_\theta$} &  {$\sigma_0$}\\
&&&{km s$^{-1}$} & {km s$^{-1}$} & {deg} & {deg} & {km s$^{-1}$} \\
\\[-10pt]\hline\\[-5pt]
 $0<l<180^\circ$  & 217 & 1.8      & 6.7      & 5.1      & 85      & 13      & 10.2\\
                        &     & $\pm0.2$ & $\pm1.6$ & $\pm1.5$ & $\pm14$ & $\pm17$ &     \\
 $30<l<180^\circ$ & 165 & 1.8      & 6.5      & 6.5      & 82      & 9       &  9.8\\
                        &     & $\pm0.2$ & $\pm1.8$ & $\pm1.7$ & $\pm16$ & $\pm15$ &     \\
 $180<l<360^\circ$& 208 & 1.8      & 5.8      & 2.0      & 78      & 154     & 10.4\\
                        &     & $\pm0.3$ & $\pm1.7$ & $\pm1.5$ & $\pm16$ & $\pm48$ &     \\
\hline
\end{tabular}
\end{minipage}
\end{table}

An analysis of the velocity field of Cepheids (Mel'nik {\it et
al.} 1999) revealed a periodic pattern along the galactic
radius-vector with a scale length of $\lambda=1.9\pm0.2$ kpc and
mean oscillation amplitudes of $f_R=6.2\pm1.2$ and
$f_\theta=2.1\pm1.2$ km s$^{-1}$. It would be interesting to see
whether the periodic pattern of the Cepheid velocity field
exhibits the same specific features as that of OB-associations.

To answer this question, we inferred the parameters of the
periodic pattern of the Cepheid velocity field in three regions
$0<l<180^\circ$, $30<l<180^\circ$, and $180<l<360^\circ$ in the
same way as we did it for OB-associations (Table 5). Note that the
exclusion of the region $0<l<30^\circ$ has no effect on the
inferred parameters of the periodic pattern in the Cepheid
velocity field. We used the parameters of circular motion and
solar-motion components that we inferred from the analysis of the
entire sample of Cepheid located within 3 kpc from the Sun [see
Table in Mel'nik {\it et al.}  (1999)]: $u_0^*=-7.6\pm0.8$ km
s$^{-1}$; $v_0=11.6\pm1.0$ km s$^{-1}$; $\Omega_0'=-5.1\pm0.2$ km
s$^{-1}$ kpc$^{-2}$; $\Omega_0''=1.0\pm0.2$ km s$^{-1}$
kpc$^{-3}$, and $\Omega=29\pm1$ km s$^{-1}$ kpc$^{-1}$. These
parameters agree within the errors with the corresponding
parameters of the velocity field of OB-associations (Table 1).

It is evident from Table 5 that in the region $30<l<180^\circ$
both radial and azimuthal velocity fields contain a periodic
component. The amplitudes of velocity variations are equal to
$f_R=f_\theta=6.5\pm1.8$ km s$^{-1}$. Numerical simulations show
that the hypothesis about the accidental nature of the periodic
variations found in the fields of radial and azimuthal residual
velocities in the region $30<l<180^\circ$ can be rejected at a
confidence level of $1-P>99\%$. The periodic pattern revealed in
the radial-velocity field in the region $180<l<360^\circ$ with an
amplitude of $f_R=5.8\pm1.7$ km s$^{-1}$ cannot be also due to
random fluctuations ($1-P>99\%$), however the variations of
azimuthal velocities with an amplitude of $f_\theta=2.0\pm1.5$ km
s$^{-1}$ can well ($P=25\%$) be due to random fluctuations.

Figure 6b shows the field of Cepheid residual velocities and,
based on   $\lambda$, $f_R$, $f_\theta$, $\varphi_R$, and
$\varphi_\theta$ from Table 5, the circular arcs corresponding to
the minima of mean radial ($V_R$) (solid line) and azimuthal
($V_\theta$) (dashed line) velocities. These arcs determine the
kinematical positions of the Cygnus, Perseus-Cassiopeia, and
Carina-Centaurus arms. The radii of arcs in Fig. 6b are determined
with an accuracy of $0.1-0.2$ kpc.

The fields of residual velocities of OB-associations and Cepheids
can be seen to have many features in common. First, the
galacticentric distances of the Carina-Centaurus ($R=6.3-6.5$ kpc)
and Perseus-Cassiopeia ($R=8.1-8.2$ kpc) arms as inferred from
analyses of radial velocities of Cepheids and OB-associations
agree well with each other. Second, both Cepheids and
OB-associations located in the quadrants I and II exhibit periodic
variations of the azimuthal velocity with an amplitude of
$f_\theta\approx6\pm2$ km s$^{-1}$, whereas the azimuthal
velocities of both populations show no periodic pattern in
quadrants III and IV. However, the velocity field of Cepheids
somewhat differs from that OB associations. Thus Cepheids do not
show the 0.3 kpc shift between the positions of the Cygnus and
Carina arms as inferred from radial velocities of OB associations.
The Galactocentric distances of the Cygnus arm as inferred from
analyses of radial velocities of Cepheids ($R=6.3\pm0.1$ kpc) and
OB-associations ($R=6.8\pm0.1$ kpc) differ significantly from each
other. In contrast to OB associations, Cepheids do not show such a
good agreement between the positions of the Cygnus and
Perseus-Cassiopeia arms as inferred from analyses of the radial
and azimuthal velocity fields.

\section*{\rm \normalsize CONCLUSION}

An analysis of the field of space velocities of OB associations
located within 3 kpc from the Sun revealed periodic variations in
the magnitude and direction of radial residual velocities $V_R$
along the galactic radius-vector with a typical scale length of
$\lambda=2.0\pm0.2$ kpc and a mean amplitude of $f_R=7\pm1$ km
s$^{-1}$.

We revealed  five kinematically distinct star-forming regions
where almost all OB-associations have the same direction of radial
residual velocity $V_R$. The radial velocities of OB associations
in the Carina-Centaurus, Cygnus, and Perseus-Cassiopeia regions
are directed mainly toward the Galactic center, whereas those of
the Sagittarius-Scutum complex and of a part of the Local system,
are directed away from the Galactic center. It is the alternation
of star-forming regions with positive and negative radial
velocities $V_R$ that determines the periodic pattern of the
radial velocity field of OB associations.

The fact that rich Carina-Centaurus and Perseus-Casiopeia
stellar-gas complexes lie in the vicinity of the minima in the
distribution of radial velocities of OB-associations indicates
that the region considered is located inside the corotation
radius. The enhanced density of high-luminosity stars in these
regions cannot be due to observational selection, because the
Carina-Centaurus and Perseus-Cassiopeia complexes are the most
distant ones among those considered in this paper (Fig. 2 and
Table 3). Furthermore, these regions cover sky areas extending
for several tens of degrees and at a heliocentric distance of 2
kpc the mean extinction averaged over such large sectors depends
little on the direction within the galactic plane. There is no
doubt, the enhanced density of high-luminosity stars in the
Carina-Centaurus and Perseus-Cassiopeia complexes is real and not
due to the extremely low extinction along the corresponding lines
of sight.

The fact that the Perseus-Cassiopeia complex ($R=8.4$ kpc) is
located inside the corotation radius imposes a lower limit on the
corotation radius, $\mid R_c-R_0\mid > 1.3$ kpc and an upper
limit on the spiral pattern speed, $\Omega_p<25$ km s$^{-1}$
kpc$^{-1}$.

Our conclusion about the corotation radius $R_c$ being located in
the outer part of the Galaxy beyond the Perseus-Cassiopeia arm is
in conflict with the conclusions of Mishurov and Zenina (1999a,
1999b) who argue that the Sun is located near the corotation
radius $\mid R_0-R_c\mid < 1$ kpc.   The latter authors based
their conclusions on the small value of radial ($f_R = 2\pm1$ km
s$^{-1}$) and large value of azimuthal ($f_\theta = 8\pm1$ km
s$^{-1}$) velocity perturbation amplitudes they inferred (Mishurov
and Zenina 1999b). Our analysis yielded a reversed amplitude
proportion with $f_R = 7\pm 1$ and $f_\theta =2\pm1$ km s$^{-1}$
both for Cepheid (Mel'nik {\it et al.} 1999) and OB-association
data (Table 1). At the same time, the pitch angle found by
Mishurov and Zenina (1999a, 1999b) for a two-armed spiral pattern
corresponds to $\lambda=2$ kpc and is consistent with our results.

As for the periodic pattern in the field of azimuthal velocities
of OB-associations, it is observed in the quadrants I and II and
is absent in the quadrants III and IV. The mean amplitude of
azimuthal velocity variations in the region of the Cygnus and
Perseus arms ($30<l<180^\circ$) is as high as $f_\theta = 6\pm2$
km s$^{-1}$, i.e., triple that of the entire sample of OB
associations. Another specific feature is the striking agreement
between the positions of the Cygnus and Perseus arms as inferred
from separate analyses of radial and azimuthal velocity fields of
OB associations. This feature can be explained by shocks that
develop when a density wave propagates through gas at a supersonic
velocity.

The periodic patterns in the residual velocity fields of Cepheids
and OB-associations have very much in common: similar scale length
of radial velocity variations along the galactic radius-vector,
$\lambda=2\pm0.2$ kpc and similar amplitudes of velocity
variations. Moreover, in both cases the azimuthal velocity fields
of objects located in  quadrants I and II exhibit a periodic
pattern, whereas no such pattern can be seen in the azimuthal
velocity field in the quadrants III and IV.

The kinematical positions of the Carina ($R = 6.3-6.5$ kpc) and
Perseus ($R = 8.1-8.2$ kpc) arms can be confidently inferred from
an analysis of the field of radial velocities of both OB
associations and Cepheids. It is the distance between these two
arm fragments that determines the scale length of the variations
of the radial velocity component along the galactic radius-vector,
$\lambda=2$ kpc.

The wavelength value that we inferred, $\lambda=2$ kpc, seems to
coincide with that of the most unstable mode of galactic disk
oscillations at the given Galactocentric distance. The very
existence of the spiral pattern suggests that the galactic disk is
marginally unstable at the solar Galactocentric distance.
Presently, the gaseous component of the galactic disk is
considered to be instrumental in maintaining such a marginal
instability (Jog and Solomon 1984; Bertin and Romeo 1988; Bertin
{\it et al.} 1989). Jog and Solomon (1984) showed that the
wavelength of the most unstable disk mode depends on the gas
fraction. They found that the wavelengths of the most unstable
disk modes at the solar Galactocentric distance should lie in the
$1-5$ kpc interval, where 1 and 5 kpc correspond to a purely
gaseous and purely stellar disk, respectively. Our result
$\lambda=2$ kpc suggests that both components play important part
in the dynamics of our Galaxy.

\subsection*{\rm \normalsize ACKNOWLEDGMENTS}

We are grateful to Yu.N.Efremov, A.V.Zasov, and A.V.Khoperskov for
the discussions, as well as for the useful remarks and advice.

\subsection*{\rm \normalsize REFERENCES}

\noindent Ambartsumian, V.A. { \it Astron. Zh.}, 1949, v. 26, p. 3
(rus).

\noindent{Barbier-Brossat, M., Figon, P.  { \it Astron. Astrophys.
Suppl. Ser.}, 2000, v. 142, p. 217.}

\noindent{Berdnikov, L.N. { \it Perem. Zvezdy}, 1987, v. 22, p.
549.}

\noindent{Berdnikov, L.N., Chernin, A.D. { \it Astron. Lett.},
1999, v. 25, p. 591}

\noindent{Berdnikov, L.N., Efremov, Yu.N. { \it Astron. Tsirk.},
1985, No. 1388, p.1.}

\noindent{Berdnikov, L.N., Dambis, A.K., Vozyakova, O.V. { \it
Astron. Astrophys. Suppl. Ser.}, 2000, v. 143, p. 211.}

\noindent{Bertin, G., Romeo, A.B. { \it Astron. and Astrophys.},
1988, v. 195, p. 105.}

\noindent{Bertin, G., Lin, C.C., Lowe, S.A., Thurstans, R.P.{ \it
Astrophys. J.}, 1989, v. 338, p. 78.}

\noindent{Blaha, C., Humphreys, R.M. { \it  Astron. J.}, 1989, v.
98, p. 1598.}

\noindent{Burton, W.B. { \it  Astron. and Astrophys.}, 1971, v.
10, p. 76.}

\noindent{Burton, W.B., Bania, T.M. { \it  Astron. and
Astrophys.}, 1974, v. 33, p. 425.}

\noindent{Creze, M., Mennessier, M.O.{ \it  Astron. and
Astrophys.}, 1973, v. 27, p. 281.}

\noindent{Dambis, A.K., Mel'nik, A.M., Rastorguev, A.S. { \it
Astron. Lett.}, 1995, v. 21, p. 291.}

\noindent{Dambis, A.K., Mel'nik, A.M., Rastorguev, A.S. { \it
Astron. Lett.}, 2001, v. 27, p. 58.}

\noindent{de Zeeuw, P. T., Hoogerwerf, R., de Bruijne, J. H. J.,
Brown, A. G. A., Blaauw, A. { \it  Astron. J.}, 1999, v. 117, p.
354.}

\noindent{Efremov, Yu.N. { \it  Astron. Astrophys. Trans.}, 1998,
v. 15, p. 3.}

\noindent{Efremov, Yu.N.,{ \it Ochagi zvezdoobrazovaniya v
galaktikakh} (Sites of Star Formation in Galaxies), Moscow: Nauka,
1989 (rus).}

\noindent{Efremov, Yu.N., Sitnik, T.G. { \it Sv. Astron. Lett.},
1988, v. 14, p.347.}

\noindent{Elmegreen, B.G.,{ \it Astrophys. J.}, 1979, v. 231, p.
372.}

\noindent{Elmegreen, B.G., Wang, M.,  { \it  Molecular Clouds in
the Milky Way and External Galaxies. Lecture Notes in Physics, v.
315}, Eds. Dickman R.L., Snell R.L., Young J.S. Amherst,
Massachusetts: Springer-Verlag, 1987, p. 240.}

\noindent{Fenkart, R.P., Binggeli, B. { \it Astron. Astrophys.
Suppl. Ser.}, 1979, v. 35, p. 271.}

\noindent{Garmany, C.D., Stencel, R.E.{ \it  Astron. Astrophys.
Suppl. Ser.}, 1992, v. 94, p. 211.}

\noindent{Gerasimenko, T.P. { \it Astron. Reports}, 1993, v. 37,
p. 480.}

\noindent{Glushkova, E.V., Dambis, A.K., Mel'nik, A.M.,
Rastorguev, A.S. { \it Astron. Astrophys.}, 1998, v. 329, p. 514.}

\noindent{Humphreys, R.M.,{ \it The Large Scale Characteristics of
the Galaxy, IAU Symp. no. 84} Ed. Burton W.B. Dordrecht: Reidel
Publ. Company, 1979, p.93.}

\noindent{Jog, C.J., Solomon, P.M. { \it  Astrophys. J.}, 1984, v.
276, p.114.}

\noindent{Kulikovski\u{i}, P.G., Zvezdnaya astromomiya, Moscow:
Nauka, 1985 (rus).}

\noindent{Lin, C.C.,Yuan, C., Shu, F.H.{ \it  Astrophys. J.},
1969, v. 155, p. 721.}

\noindent{Mel'nik, A.M.,  Dambis, A.K., Rastorguev, A.S. { \it
Astron. Lett.}, 1999, v. 25, p. 518.}

\noindent{Mel'nik, A.M., Efremov, Yu.N. { \it Astron. Lett.},
1995, v. 21, p. 10.}

\noindent{Mel'nik, A.M., Sitnik T.G., Dambis, A.K., Efremov,
Yu.N., Rastorguev, A.S. { \it Astron. Lett.}, 1998, v. 24, p.
594.}

\noindent{Mishurov, Yu.N., Pavlovskaya, E.D., Suchkov, A.A.{ \it
Sv. Astronony}, 1979, v.23, p.147.}

\noindent{Mishurov, Yu.N., Zenina, I.A., { \it Astron.
Astrophys.}, 1999a, v. 341, p. 81.}

\noindent{Mishurov, Yu.N., Zenina, I.A.,{ \it  Astron. Reports},
1999b, v. 43, p. 487.}

\noindent{Mishurov, Yu.N., Zenina, I.A., Dambis, A.K., Mel'nik,
A.M., Rastorguev, A.S.{ \it Astron. Astrophys.}, 1997, v. 323, p.
775.}

\noindent{Morgan, W.W., Sharpless, S., Osterbrock, D. { \it
Astron. J.}, 1952, v. 57, p. 3.}

\noindent{Press, W.H., Flannery, B.P., Teukoisky, S.A.,
Vetterling, W.T.) { \it Numerical Recipes: The art of scientific
computing.} Cambridge: Cambridge Univ. Press, 1987.}

\noindent{Rastorguev, A.S., Durlevich, O.V., Pavlovskaya, E.D.,
Filippova, A.A.,{ \it Astron. Lett.}, 1994, v. 20, p. 591.}

\noindent{Roberts, W.W., { \it Astrophys. J.}, 1969, v. 158, p.
123.}

\noindent{Sitnik, T.G., Mel'nik, A.M., { \it Astron. Lett.}, 1996,
v. v. 22, p. 422.}

\noindent{Sitnik, T.G., Mel'nik, A.M., { \it Astron. Lett.}, 1999,
v. 25, p. 156.}

\noindent{Sitnik, T.G., Mel'nik, A.M., Pravdikova, V.V. { \it
Astron. Reports}, 2001, v. 45, p. 34.}

\noindent{The Hipparcos and Tycho Catalogues. European Space
Agency, 1997, v. 1--20.}

\clearpage
\thispagestyle{empty}
\begin{table}[t]
\begin{minipage}{8 cm}
 \footnotesize
 \tablename{ 2. Residual velocities of OB associations \hfill}
\begin{tabular}{lcccccccccc}
 \\[-5pt]\hline\\ [-5pt]
{Association} &{$l$} & {$b$} & {$r$} & {$R$} &
{$V_R$} & {$V_\theta$}  &  {$V_z,$} & {$n_r$}& {$n_l$}& {$N$} \\
& & & {kpc} & {kpc} & {km s$^{-1}$} & {km s$^{-1}$} & {km s$^{-1}$} & & & \\
[-0 pt] \hline\\[-5 pt]
Sgr OB5   &0.$^\circ$0  &-1.$^\circ$2  &2.4 &4.7&7.5  &9.0&8.0  &2  &3 &31\\
Sgr OB1   &7.6  &-0.8  &1.3 &5.8 &7.3  &-5.4 &1.0  &37 &29 &66\\
Sgr OB7  &10.7  &-1.6  &1.4 &5.7 &9.2  &3.9  &-14.4&3  &2  &4\\
Sgr OB4  &12.1  &-1.0  &1.9 &5.2 &4.8  &-0.0 &-0.1 &9  &3  &15 \\
Ser OB1  &16.7  &0.1   &1.5 &5.7 &13.0 &-0.9 &1.4  &17 &12 &43\\
Sct OB3  &17.3  &-0.7  &1.3 &5.8 &2.7  &1.7  &3.0  &8  &3  &10\\
Ser OB2  &18.2  &1.6   &1.6 &5.6 &13.7 &-1.9 &-3.7 &7  &5  &18\\
Sct OB2  &23.2  &-0.5  &1.6 &5.7 &26.3 &-2.3 &4.9  &6  &6  &13\\
Vul OB1  &60.4  &0.0   &1.6 &6.5 &6.4  &-3.3 &6.1  &9  &8  &28\\
Vul OB4  &60.6  &-1.2  &0.8 &6.7 &2.7  &-0.1 &-0.1 &3  &3  &9\\
Cyg OB3  &72.8  &2.0   &1.8 &6.8 &-14.5&-7.8 &-4.0 &30 &18 &42\\
Cyg OB1  &75.8  &1.1   &1.5 &6.9 &-3.9 &-8.3 &1.2  &34 &14 &71\\
Cyg OB9  &77.8  &1.8   &1.0 &7.0 &-5.8 &-12.2&0.2  &10 &8  &32\\
Cyg OB8  &77.9  &3.4   &1.8 &7.0 &-3.9 &-13.9&12.8 &9  &10 &21\\
Cyg OB4  &82.7  &-7.5  &0.8 &7.0 &15.5 &4.8  &2.7  &2  &2  &2\\
Cyg OB7  &89.0  &0.0   &0.6 &7.1 &2.5  &2.4  &3.7  &21 &28 &29\\
Lac OB1  &96.7  &-17.7 &0.5 &7.2 &-3.6 &-2.8 &1.6  &2  &2  &2\\
Cep OB2 &102.1  &4.6   &0.7 &7.3 &-3.2 &-0.1 &2.8  &37 &47 &59\\
Cep OB1 &104.2  &-1.0  &2.8 &8.2 &-8.7 &-14.3&3.1  &17 &24 &58 \\
Cas OB2 &112.0  &0.0   &2.1 &8.1 &-19.0&-3.3 &6.2  &7  &5  &41\\
Cep OB3 &110.4  &2.6   &0.7 &7.4 &-3.9 &-4.9 &0.5  &18 &15 &26\\
Cas OB5 &116.1  &-0.5  &2.0 &8.2 &-12.7&-2.7 &-11.2&16 &13 &52\\
Cas OB4 &120.1  &-0.3  &2.3 &8.5 &3.4  &0.8  &-6.4 &7  &7  &27\\
Cas OB14&120.4  &0.7   &0.9 &7.6 &11.6 &-2.7 &2.3  &4  &3  &8\\
Cas OB7 &123.0  &1.2   &2.0 &8.4 &-10.3&-9.2 &-2.7 &4  &8  &39\\
Cas OB1 &124.7  &-1.7  &2.0 &8.4 &-6.8 &-2.5 &-7.0 &5  &3  &11 \\
NGC 457 &126.7  &-4.4  &2.0 &8.4 &-1.7 &1.0  &-6.7 &4  &2  &4\\
Cas OB8 &129.2  &-1.1  &2.3 &8.7 &2.3  &2.5  &-3.7 &14 &9  &43\\
Per OB1 &134.7  &-3.2  &1.8 &8.5 &-7.8 &-12.2&-5.5 &81 &63 &167\\
Cas OB6 &135.0  &0.8   &1.8 &8.4 &-11.2&-6.8 &-3.8 &12 &13 &46\\
Cam OB1 &141.1  &0.9   &0.8 &7.7 &-0.4 &6.2  &0.3  &30 &33 &50 \\
Cam OB3 &147.0  &2.8   &2.6 &9.4 &9.5  &-17.6&16.5 &3  &3  &8\\
Per OB2 &160.3  &-16.5 &0.3 &7.4 &16.0 &9.6  &-0.9 &7  &7  &7\\
Aur OB1 &173.9  &0.1   &1.1 &8.2 &-4.9 &0.8  &-2.7 &26 &20 &36\\
Ori OB1 &206.9  &-17.7 &0.4 &7.4 &8.2  &0.5  &0.3  &62 &59 &70\\
Aur OB2 &173.3  &-0.2  &2.4 &9.5 &-3.5 &12.0 &-2.8 &4  &2  &20\\
Gem OB1 &189.0  &2.2   &1.2 &8.3 &2.2  &5.2  &3.9  &18 &17 &40\\
Mon OB1 &202.1  &1.1   &0.6 &7.6 &6.3  &0.5  &1.8  &7  &7  &7\\
Mon OB2 &207.5  &-1.6  &1.2 &8.2 &-0.4 &10.9 &-0.5 &26 &18 &32\\
CMa OB1 &224.6  &-1.6  &1.1 &7.9 &7.2  &3.5  &-7.0 &8  &10 &17\\
Coll 121&238.5  &-8.4  &0.6 &7.4 &5.9  &-4.1 &0.1  &10 &13 &13\\
NGC 2362&237.9  &-5.9  &1.2 &7.8 &0.5  &3.8  &4.5  &6  &3  &9\\
Coll 140&244.5  &-7.3  &0.3 &7.2 &-2.7 &7.6  &-1.5 &5  &6  &6\\
Vela OB2&262.1  &-8.5  &0.4 &7.2 &2.2  &-8.7 &0.8  &13 &12 &13\\
Vela OB1&264.9  &-1.4  &1.5 &7.4 &-0.5 &-1.9 &-2.5 &18 &18 &46 \\
Car OB1 &286.5  &-0.5  &2.0 &6.8 &0.8  &3.4  &-2.2 &39 &18 &126 \\
Tr 16   &287.3  &-0.3  &2.1 &6.8 &-2.3 &-1.8 &-2.6 &5  &2  &18\\
Coll 228&287.6  &-1.0  &2.0 &6.8 &-6.6 &10.0 &-2.7 &9  &2  &15 \\
Car OB2 &290.4  &0.1   &1.8 &6.7 &-4.3 &2.2  &0.3  &22 &12 &59\\
Cru OB1 &294.9  &-1.1  &2.0 &6.5 &-13.5&-6.1 &-0.2 &33 &19 &76\\
NGC 3766&294.1  &-0.0  &1.5 &6.6 &-7.3 &7.9  &0.5  &2  &2  &12\\
Cen OB1 &304.2  &1.4   &1.9 &6.2 &-16.8&0.6  &-2.4 &32 &32 &103\\
Hogg 16 &307.5  &1.4   &1.5 &6.3 &-7.4 &19.9 &-11.7&3  &3  &5\\
NGC 5606&314.9  &1.0   &1.5 &6.1 &3.2  &12.4 &-6.6 &3  &2  &5\\
Ara OB1A&337.7  &-0.9  &1.1 &6.1 &15.4 &10.0 &-3.8 &8  &10 &53\\
Sco OB1 &343.7  &1.4   &1.5 &5.6 &6.0  &4.2  &1.1  &28 &16 &76\\
Sco OB2 &351.3  &19.0  &0.1 &7.0 &-3.3 &-3.1 &0.7  &10 &10 &10\\
HD 156154&351.3 &1.4   &2.1 &5.0 &-15.6&-5.1 &-2.1 &3  &2  &4 \\
Sco OB4 &352.7  &3.2   &1.0 &6.2 &-15.0&2.7  &-3.2 &7  &4  &11\\
\hline
\end{tabular}
\end{minipage}
\end{table}

\clearpage
\begin{figure}
 \includegraphics{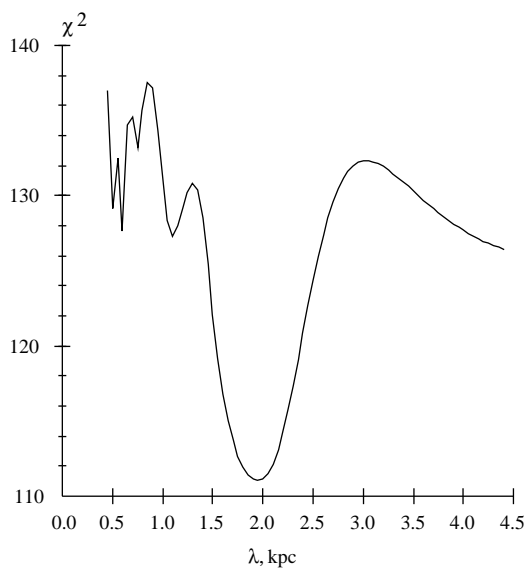} \vspace{6.0 cm}   \caption{\footnotesize The
function $\chi^2(\lambda)$ for the joint solution of the system of
equations (3) and (4) for the entire sample of OB associations
\hfill}
\end{figure}

\newpage
\begin{figure}
\includegraphics{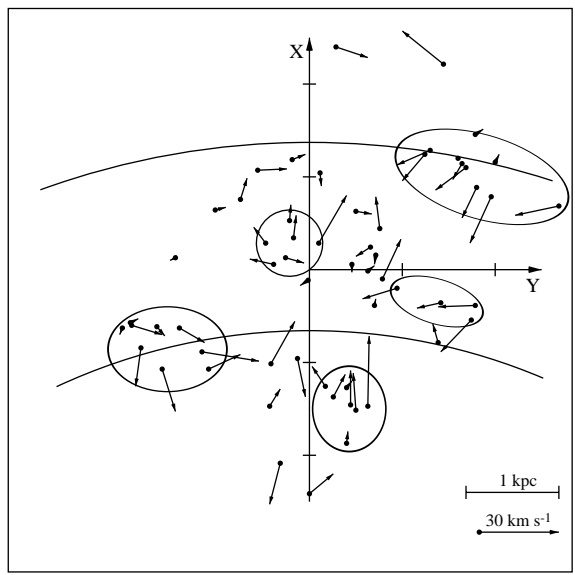} \vspace{8 cm} \caption{\footnotesize The field of space
velocities of OB-associations projected onto the galactic plane.
The $X$-axis is directed away from the Galactic center, and the
$Y$-axis is in the direction of galactic rotation. The Sun is at
the origin. The circular arcs correspond to the maximum radial
component $V_R$ of residual velocity toward the galactic center.
One can see five star-forming regions where almost all
OB-associations have the same direction of radial velocity $V_R$.
\hfill}
\end{figure}

\newpage
\begin{figure}
\includegraphics{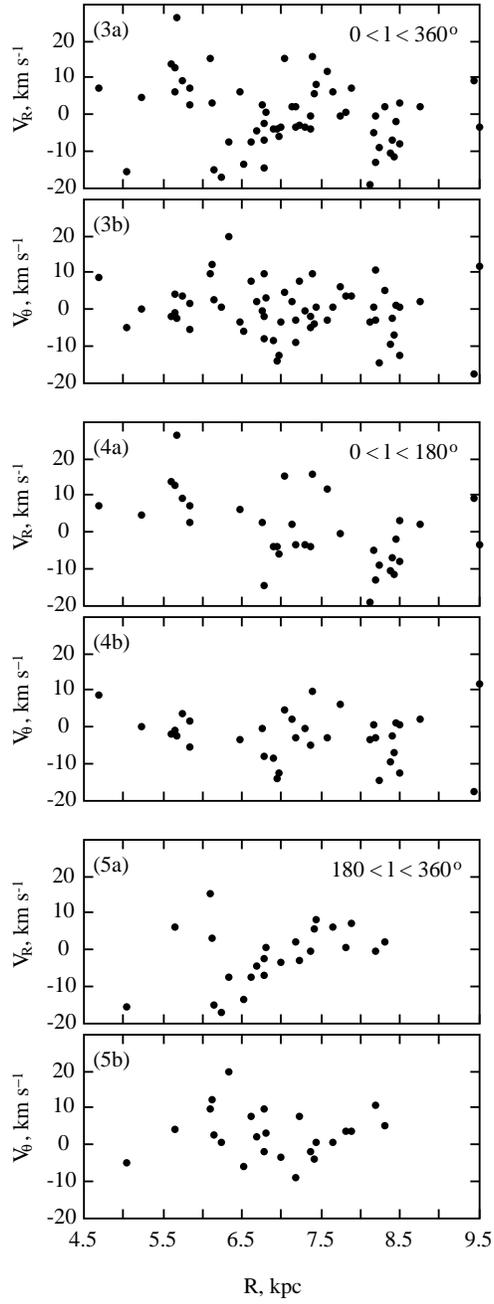} \vspace{16 cm} \caption{4, 5. \footnotesize Residual
velocities $V_R$ and $V_\theta$ of OB-associations as a function
of Galactocentric distance $R$. The whole sample and  the regions
$0<l<180^\circ$ and $180<l<360^\circ$ are concidered.}
\end{figure}

\addtocounter{figure} {2}

\newpage
\begin{figure}
\includegraphics{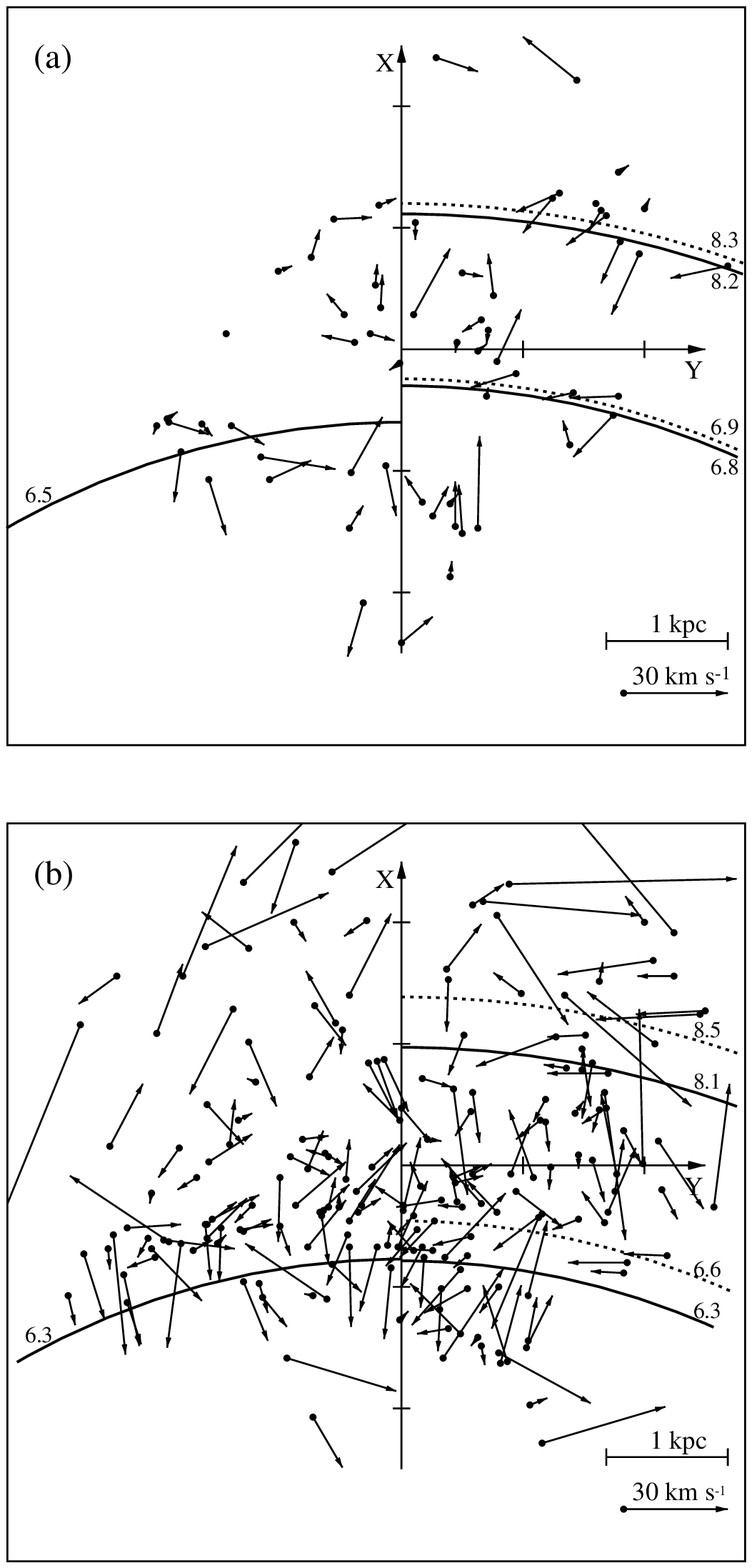} \vspace{14 cm} \caption{ \footnotesize The field of
residual velocities of (a) OB-associations and (b) Cepheids. The
circular arcs correspond to the positions of the arm fragments as
inferred from analyses of radial (solid line) and azimuthal
(dashed line) residual velocities in the regions $30<l<180^\circ$
and $180<l<360^\circ$.}
\end{figure}

\end {document}